\documentclass[a4paper]{article}
\usepackage{multirow}
\usepackage{INTERSPEECH2022}
\usepackage{fourier} 
\usepackage{array}

\usepackage{makecell,url}

\usepackage{changepage}

\makeatletter
\newcommand\footnoteref[1]{\protected@xdef\@thefnmark{\ref{#1}}\@footnotemark}
\makeatother

\usepackage{xcolor}

\title{The Unreliability of Acoustic Systems in Alzheimer's Speech Datasets with Heterogeneous Recording Conditions}
\name{Lara Gauder$^{1,2}$, Pablo Riera$^1$, Andrea Slachevsky$^3$, Gonzalo Forno$^3$,\\Adolfo M. García$^4$, Luciana Ferrer$^1$}
\address{\vspace{-0.2cm}
    $^1$Instituto de Investigación en Ciencias de la Computación, UBA-CONICET, Argentina\\
    $^2$Departamento de Computación, Facultad de Ciencias Exactas y Naturales, UBA, Argentina \\  
    $^3$Facultad de Medicina , Universidad de Chile, Chile \\
    $^4$Centro de Neurociencias Cognitivas, Universidad de San Andrés, Argentina
}

\email{\{mgauder,priera,lferrer\}@dc.uba.ar, adolfo.garcia@gbhi.org}

\begin{document}

\maketitle
\begin{abstract}
Automated speech analysis is a thriving approach to detect early markers of Alzheimer’s disease (AD). Yet, recording conditions in most AD datasets are heterogeneous, with patients and controls often evaluated in different acoustic settings. While this is not a problem for analyses based on speech transcription or features obtained from manual alignment, it does cast serious doubts on the validity of acoustic features, which are strongly influenced by acquisition conditions. We examined this issue in the ADreSS$_o$ dataset, derived from the widely used Pitt corpus. We show that systems based on two acoustic features, MFCCs and Wav2vec 2.0 embeddings, can discriminate AD patients from controls with above-chance performance when using only the non-speech part of the audio signals. We replicated this finding in a separate dataset of Spanish speakers. Thus, in these datasets, the class can be partly predicted by recording conditions. Our results are a warning against the use of acoustic systems for identifying patients based on non-standardized recordings. We propose that acoustically heterogeneous datasets for dementia studies should be either (a) analyzed using only transcripts or other features derived from manual annotations, or (b) replaced by datasets collected with strictly controlled acoustic conditions.
\end{abstract}

\section{Introduction}
\vspace{-0.1cm}

Alzheimer’s disease (AD) is a chronic neurodegenerative disorder characterized by a progressive decline of cognitive and functional abilities \cite{american2013diagnostic}, often including oral communication deficits, even at early stages \cite{ross1990speech, GARCIA202266, garcia2022unusual}. Thus, many studies rely on speech processing approaches for automatic detection of the disease \cite{pulido2020alzheimer, lopez2018advances, lopez2013}. Promisingly, speech evaluation systems could be used by doctors as diagnostic-support tools, providing a potentially robust and replicable measurement based on the patient's speech.

Systems for automatic AD detection from speech use a variety of approaches. Several studies use features extracted from speech transcripts \cite{orimaye2017predicting, searle20_interspeech, sanz2022automated}, while others use acoustic features extracted directly from the speech signal \cite{meilan2014speech, adresso_gauder21}, or combine both sources of information \cite{fraser2016linguistic, martinc20_interspeech}. The acoustic features used for this task include statistics derived from mel frequency cepstral coefficients (MFCCs) \cite{fraser2016linguistic, hernandez2018, meghanani2021}, and other acoustic features like F0, energy, and spectrograms \cite{haider2019assessment, nasreen21_interspeech}. Some recent works use speech-based embeddings extracted from deep neural networks (DNNs). In \cite{perez2021influence} and \cite{pappagari20_interspeech} authors use embeddings extracted from DNNs trained for speaker identification. In our recent work \cite{adresso_gauder21}, we explored the use of wav2vec 2.0 \cite{baevski2020wav2vec} and trill embeddings \cite{shor2020towards}, both trained with self-supervision.    

Most of the studies mentioned above used data from the DementiaBank’s Pitt corpus \cite{PittCorpus}, collected in the 1980s. Both the ADreSS and ADreSS$_o$ datasets used in the respective challenges \cite{LuzHaiderEtAl20ADReSS, luz21_interspeech} were extracted from this corpus. Acoustic conditions in this dataset are markedly heterogeneous, arguably because data were not collected with the goal of performing automatic speech analyses. The recorded audio samples include variable noises, microphone movements, additional voices, and different bit rate encoding methods, among other issues. 
Since the acoustic features being used in the literature may encode not only information from participants but also from the recording conditions, we deem it essential to examine whether acquisition conditions are biasing reported results. For example, if the signals' acoustic condition were somehow correlated with the class (e.g., because AD patients were recorded in a different room or with a different device than control patients), systems or analyses using acoustic features could yield spurious results, possibly inflating the apparent predictive power of these features. 

We studied this issue on the ADreSS$_o$ dataset and a proprietary dataset collected in Chile. Specifically, we analyzed the classification performance of two systems based on acoustic features, namely, MFCCs and Wav2vec 2.0 embeddings, extracted from speech or non-speech regions. We find that, on both datasets, systems that use only the non-speech part of the signals perform above chance, indicating that the class of the samples (AD or control) can be partly predicted from the recordings’ acoustic characteristics. 
This worrisome finding suggests that acoustic features should not be used for automatic prediction or for statistical analysis when the dataset's recording conditions were not carefully controlled during data collection.

\vspace{-0.2cm}
\section{Datasets}
\vspace{-0.1cm}
Two AD datasets are used in this paper, both composed of audio recordings of participants describing the Cookie Theft picture from the Boston Diagnostic Aphasia Exam. Below we describe the specifications for each dataset, as well as the pre-processing variants explored. 

\vspace{-0.1cm}
\subsection{ADreSS$_o$ Dataset}
\label{sec:adresso}
\vspace{-0.1cm}
The Alzheimer's Dementia Recognition through Spontaneous Speech \textit{only} (ADreSS$_o$) challenge was organized at Interspeech 2021 \cite{luz21_interspeech}. In this work we use the dataset released for training purposes during the challenge, since the labels for the test data were not released. The training set consists of one recording for each of 77 control subjects and 81 AD subjects. Each participant recorded a single session. The mean recording duration is $72.73$ seconds (stdev $26.75$). The dataset is balanced by sex and age. 

Since recordings include both the speech from the participant and the experimenter, the dataset includes segmentation information indicating where each of the them speaks. In our initial inspection of the data, though, we found that this information was inaccurate for many of the audio files.  In \cite{adresso_gauder21}, using only the audio files for which the manual segmentation had no obvious issues, we showed that the effect of including the experimenter speech appears to be negligible. Further, the main conclusions from this paper are drawn from the analysis of the non-speech parts of the signals and, hence, are unaffected by this issue. For these reasons, in this work, we do not use the provided segmentation information.

We show results for four versions of this dataset. The first version is composed of the waveforms released during the challenge which were acoustically enhanced by the organizers with stationary noise removal, and audio volume normalization \cite{luz21_interspeech}. The second version of this dataset is composed of the original recordings before enhancement. These waveforms were provided by the organizers of the challenge upon request. Finally, we evaluate performance on two more versions of the dataset obtained after applying enhancement approaches (described in Section \ref{sec:speech_enhancement}) on the original recordings. 

The ADreSS$_o$ waveforms have a sampling rate of 44.1 kHz, except for two cases which have 8 kHz sampling rate and which we discarded for our analysis. We discarded one more sample where the speech was not audible in the original waveform. 
The original recordings are coded in mp3 format and their bit rate is constant for 16 of the control samples, and variable for all other control and all AD samples. All waveforms were downsampled to 16kHz to allow for the use of pre-trained models (see Section \ref{sec:methods}).

\vspace{-0.1cm}
\subsection{Spanish Alzheimer's Disease Dataset} \label{sec:chile-dataset}
\vspace{-0.1cm}
The Spanish Alzheimer's Disease dataset (SpanishAD, for short) \cite{diaz2022multidimensional, birba2021multimodal, sanz2022automated} is composed of 39 Spanish-speaking subjects, where 18 are control subjects and 21 have been clinically diagnosed with AD by expert neurologists.
The dataset was collected at the Memory and Neuropsychiatry Clinic hosted by Universidad de Chile and Hospital del Salvador, Chile. The recordings -- as in ADreSS$_o$, one per speaker -- have a mean duration of $84.77$ seconds (stdev $27.98$). We discarded four females from the AD group and one male from the control group to obtain the same proportion of males (76\%) in both groups. Age distribution is similar across groups.

Recording were made in several formats, listed in Table~\ref{tab:codec}. The quality of the recordings varies widely due to the different codecs and sampling rates used, as well as to different background noises and reverb characteristics. We down-sampled all waveforms to 11025 Hz, to homogenize the bandwidths, and then up-sampled them to 16 kHz. 
Each recording was manually aligned to transcriptions by an annotator at IPU (inter pause unit) level. An IPU is defined as consecutive regions of speech with no more than 0.2 seconds of pause between them. Coughs, laughs and filled pauses, considered speech in our work, are also carefully annotated and aligned. Each IPU was also labeled with the speaker information. 
We show results for three different versions of this dataset, the original one and two enhanced versions (see Section \ref{sec:speech_enhancement}).

\begin{table}[t]
    \centering
        \footnotesize
    \begin{tabular}{|c|c|c|c|}
        \hline
        \thead{Pathology} & \thead{SR} & \thead{codec} & \thead{\#Samples}  \\
        \hline
        \hline
        \multirow{2}{*}{AD} & 11025 & \multirow{2}{*}{PCM} & 16 \\
        \cline{2-2}
        \cline{4-4}
        & 48000 &  & 5 \\
        \hline
        \multirow{4}{*}{CTR} & 11025 & PCM & 2 \\
        \cline{2-4}
         & \multirow{2}{*}{44100} & AAC LC & 10 \\
         \cline{3-4}
         &  & PCM & 1 \\
         \cline{2-4}
         & 48000 & PCM & 5 \\
         \hline
    \end{tabular}
    \vspace{0.05cm}
    \caption{Number of samples in the SpanishAD dataset per class, sampling rate (SR) in Hertz, and codec.\vspace{-0.2cm}}
    \label{tab:codec}
\end{table}

\vspace{-0.1cm}
\section{Methods}
\vspace{-0.1cm}
\label{sec:methods}
For our analyses, we implement two simple systems that use two different acoustic features as input and the same pre-processing steps and classifier approach. A voice-activity detection system is used to determine the speech and non-speech regions. Then, MFCC or Wav2vec 2.0 embeddings are extracted from each selected segment, either from the raw signal or an enhanced signal.  The classifier is given by a simple deep neural network. All these components are described below. Examples of the processed waveforms and the alignments can be found in \url{https://github.com/marialaraa/adresso_analysis}.

\vspace{-0.1cm}
\subsection{Speech enhancement}\label{sec:speech_enhancement}
\vspace{-0.2cm}
We implemented two speech enhancement procedures: loudness normalization and noise reduction. Since audio samples in both datasets present variable recording volumes, with the experimenter and background noises sometimes being louder than the participant, we perform loudness normalization following the EBU R128 standard \cite{ebu2011loudness}. The normalization is done over sliding windows, which results in experimenter, participant and background noises having similar loudness after processing. While this may seem undesirable, as we will see, for the SpanishAD dataset, it seems to help reduce the effect of the acoustic conditions when used in combination with noise reduction. 
Noise reduction is implemented using the FullSubNet CNN model \cite{hao2020fullsubnet}, which effectively subtracts stationary ambient noise and some  transient noises, though not reverberation. Listening tests were performed, verifying that speech characteristics were preserved but with better overall quality. Examples of enhanced waveforms can be found in the webpage listed above.

\vspace{-0.1cm}
\subsection{Voice Activity Detection}
\vspace{-0.2cm}
For voice activity detection (VAD) we used the pre-trained Silero model \cite{SileroVAD}. 
This model outputs scores extracted over windows of 0.1 seconds (without overlap) measuring the probability that speech is present on that window. Speech regions are then determined by comparing this score with a tunable threshold. After some exploration, for the purpose of this paper, we set this threshold to $0.5$, the default value. Exploring the VAD output at this threshold on a number of different samples, we found that while some background noises are sometimes labeled as speech, speech is rarely labeled as non-speech. 
Examples of the VAD output can be found in the webpage listed above.

\vspace{-1mm}
\subsection{Wav2vec 2.0 embeddings}
\vspace{-2mm}
Wav2Vec 2.0 (from now on, wav2vec2) is a framework for self-supervised learning of representations from raw audio signals~\cite{baevski2020wav2vec}. Wav2vec2 models can be fine-tuned for the task of automatic speech recognition, leading to outstanding results, showing that the model contains rich phonetic information. Furthermore, pre-trained wav2vec2 models can be used to extract low-level embeddings which have been successfully used for a variety of other tasks, including 
emotion recognition \cite{pepino21_interspeech, macary2021use, boigne2020transfer}, suggesting that these embeddings contain information beyond the phonetic content. In a recent work, we proposed to use wav2vec2 features for AD detection \cite{adresso_gauder21}, though, as we show in this work, we now know that the excellent results obtained in that work are likely partly explained by spurious correlations between the acoustic conditions of the recordings and the patient's diagnosis.

For the ADreSS$_o$ database we used the Wav2Vec 2.0 Base model\footnote{https://huggingface.co/facebook/wav2vec2-base} which provides embeddings of size 768. This model was trained using only English speech. For the SpanishAD dataset, we use the XLSR-53 model,\footnote{https://huggingface.co/facebook/wav2vec2-large-xlsr-53} which was trained with multiple languages, including Spanish. The embeddings from this model are 512-dimensional. In both cases, the models are only pre-trained in self-supervised fashion, without fine-tuning for any specific task. Both models produce embeddings every 20 milliseconds. 

In the analyses, we explore the performance of systems when the features are extracted only on certain regions of the signal, e.g., only speech or non-speech regions. Wav2vec2 embeddings are contextualized, potentially integrating information from any region in the signal in each embedding. Hence, to ensure that no information from regions not included in the analysis is used in the computation of the embeddings, we extract these embeddings by providing only the selected regions to the model (one at a time), rather than extracting them over the full signal and then selecting the regions of interest. Finally, the embeddings obtained for each region within a waveform are concatenated, normalized (subtracting the mean and dividing by the standard deviation on each dimension), and fed into the classifier. 

\vspace{-1mm}
\subsection{Mel-frequency Cepstral Coefficients}
\vspace{-2mm}
Mel-frequency Cepstral Coefficients (MFCCs) \cite{mermelstein1976distance, bridle1974} have been one of the most standard features used in speech processing for several decades. They are used for a large variety of tasks, including high-level ones like emotion recognition \cite{koolagudi2012emotion}, speaker verification \cite{bai2021speaker}, and AD detection \cite{pulido2020alzheimer}. Here, we extract MFCCs using the \textit{librosa} library \cite{mcfee2015librosa}. We extract 20 coefficients over windows of 20 miliseconds with an overlap of 10 miliseconds. 
When working with specific regions of the audio signals, we extract the features using the same process as for wav2vec2 embeddings, computing the features over individual regions of interest, concatenating the resulting vectors, and normalizing them, before feeding them to the classifier.

\vspace{-1mm}
\subsection{Classification model}
\vspace{-2mm}
For classification, we use a simple DNN with a first time-distributed dense layer to reduce the input size to 64 dimensions, followed by a 1D convolution with a kernel size of 3 and 128 output channels. Both layers use ReLu activations and batch normalization. The output of the second layer is averaged over time to obtain a single vector per sample, which is then fed into the output layer with sigmoid activation. The DNN is trained to optimize binary cross-entropy.
For computational reasons, the samples are processed in chunks, taking 5-seconds worth of features (after concatenation over all regions of interest) at a time to input to the model. Chunks overlap by one second. The final scores are then computed by averaging the scores output by the model for all chunks corresponding to an audio sample.
More details on the architecture can be found in \cite{adresso_gauder21}.

\vspace{-2mm}
\section{Results and Discussion}\label{sec:results}
\vspace{-1mm}
Systems for each dataset and configuration 
are trained and evaluated using an 8-fold cross-validation approach. To assess the variability due to the specific choice of folds, we run each system using 50 random seeds. For each seed, we obtain the area under the curve (AUC) of false positive versus false negative rates using the pooled scores from all samples. This results in 50 estimates of the AUC for each system. 
We show these results in the form of box plots for each dataset (ADreSS$_o$ and SpanishAD), each input feature (MFCC or wav2vec2), and each type of pre-processing approach (no pre-processing, ORIG; the ADreSS$_o$ challenge's pre-processing, CHALL; noise reduction only, NR; and loudnorm followed by noise reduction, LN+NR). For each of those cases, we show results for systems trained and evaluated either on speech regions (SPEECH) or non-speech regions (NON-SP). The speech and non-speech regions can be obtained by the VAD system or, in the case of SpanishAD, from the manual annotations (MANUAL). Note that speech regions include both the speech from the experimenter and the participant. In the case of the SpanishAD dataset, for which we have accurate annotations of where the participant speaks, we also show results restricting the feature extraction to regions where only the participant speaks (PAR). 

\begin{figure*}[t]
    \centering
    \includegraphics[width=17cm]{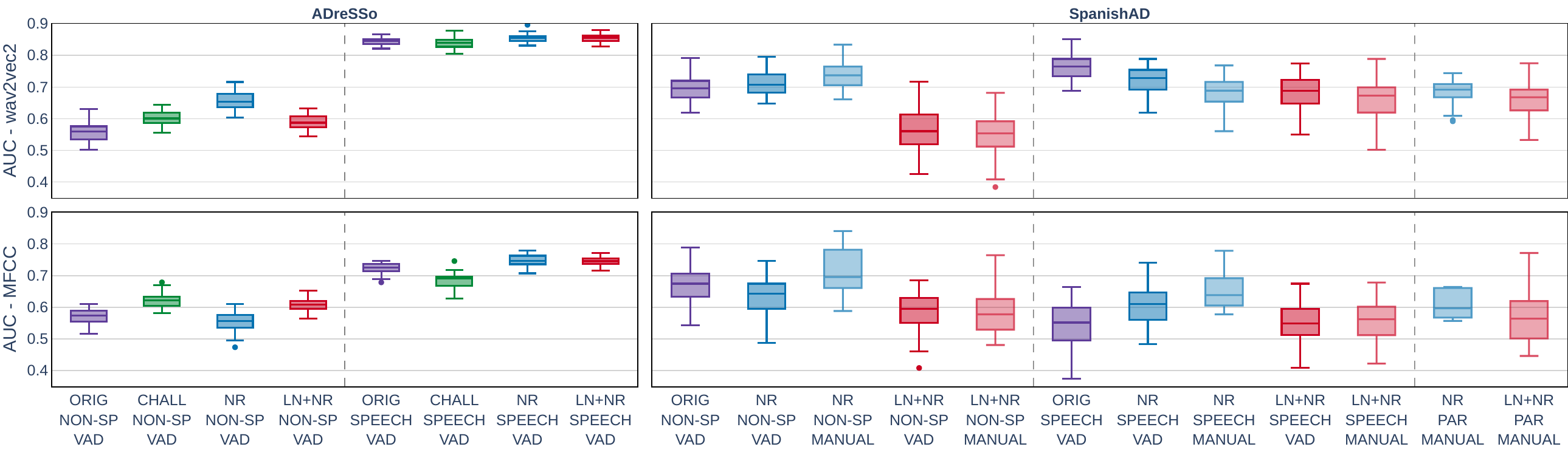}
    \vspace{-0.7cm}
    \caption{AUC on the ADreSS$_o$ and SpanishAD datasets for different features, pre-processing methods and selected regions. See Section~\ref{sec:results} for a description of the system names.}
    \label{fig:results}
\end{figure*}

The left plots in Figure \ref{fig:results} show results for the ADreSS$_o$ dataset. We can see that results over non-speech regions for both types of features are significantly better than those of a random system (which would have an AUC of 0.5), reaching a mean AUC value of 0.66. These results indicate the presence of spurious correlations between the samples' class and the acoustic conditions.
The enhancement applied by the organizers of the challenge and our own enhancement methods do not consistently help mitigate this effect. Surprisingly, the NR method increases the AUC of the wav2vec2 embeddings over non-speech regions, apparently emphasizing the spurious correlations or, perhaps, the small amount of speech that is sometimes labeled as non-speech by the VAD system. 
Notably, inspecting the average power spectrum of the ADreSS$_o$ original waveforms we found that waveforms with different bit rate mode have distinct noise floor levels. In particular, the information in the 14 to 16 kHz frequency region can be used reliably to predict the bit rate mode. 
Since, in this dataset, variable bit rate was only used on control samples (see Section \ref{sec:adresso}), the model can use this information to help predict the class of the samples explaining, at least in part, the performance obtained on non-speech regions.

To analyze this worrisome phenomenon further, we use the SpanishAD dataset for which manual alignments of IPUs annotated with the speaker identity are available. The right plots in Figure \ref{fig:results} show these results. Systems that only see non-speech regions again show above-chance performance, this time reaching a mean AUC of 0.75. Enhancement with noise reduction (NR) does not mitigate the problem, but loudness normalization followed by noise reduction (LN+NR) tends to reduce the effect, bringing performance closer to chance level for both feature types. We hypothesize that, for this dataset, LN helps the NR process better achieve its goal, since LN on its own does not have this effect (results not shown to avoid cluttering the figure).
Comparing the automatic with the manual determination of the non-speech regions, we see that results are very similar, with a slight increase in AUC for the NR case for the manual annotations where we know that no speech is included.  
Focusing now on the results over speech regions we see that in most cases, results for the same pre-processing (same-color boxes) are better when using the speech regions than when using non-speech regions, suggesting that these acoustic features contain useful information for the diagnosis. A notable exception are the MFCC systems on SpanishAD, where it appears that the features are unable to capture information about the participant's diagnosis, perhaps due to the very limited amount of training data.
Unfortunately, even when results over speech regions are better than over non-speech regions, we cannot assume that the results over speech regions would hold in absence of spurious correlations. This is true even when the results on non-speech regions are close to chance, as in the case of the LN+NR results in SpanishAD or of the original signals for ADreSS$_o$, since biases in acoustic characteristics may be more salient over speech than over non-speech regions.

While for the ADreSS$_o$ dataset the above-chance results over the non-speech regions can be, at least partly, explained by the fact that bit rate modes correlate with the class of the samples, for the SpanishAD dataset, a possible explanation can be found in differences in sampling rates and codecs. Table \ref{tab:codec} shows that most AD but only a minority of control waveforms were collected with a low sampling rate. While we attempted to mitigate the effect of this bias by downsampling all recordings to the lowest sampling rate in the dataset, the sampling rate is probably correlated with other characteristics of the recording device, which are still present in the signal after downsampling. Further, a majority of the control cases were recorded with a codec that is never used for AD participants. These acquisition characteristics can be reflected in the acoustic features potentially allowing the model to detect the class with above-chance performance using only non-speech regions.

The box plots on the right of Figure \ref{fig:results} show the results obtained only on the participant's speech (PAR), discarding the experimenter's speech. We can see that the influence of the experimenter's speech, in this dataset and for these two systems, is negligible. This is consistent with our conclusions using the ADreSS$_o$ data in \cite{adresso_gauder21}. Hence, our results over speech regions are likely not significantly affected by potential biases in the experimenter's identity or behavior.

The results presented in this section raise concerns about the use of these two datasets for studies involving acoustic features. While we show results on two specific datasets, similar problems may be present in other datasets being used for dementia studies, since such datasets are often collected for studying characteristics that are robust to the acoustic conditions, like those based on speech transcripts. Acoustic heterogeneity or biases in the recording conditions are of no concern in those cases and, hence, not necessarily controlled during data collection. Yet, when those datasets are used for studies involving acoustic features, these issues become major problems that hinder the analysis. The analysis proposed here can be used to diagnose datasets and detect clear-cut cases in which acoustic features should not be used. 

Much of the work done in dementia detection from speech is based on speech rate or rhythm estimates and on features derived from pitch signals. Unfortunately, the analysis in this work cannot be done using those types of features since they are only defined over speech regions. Yet, we would like to note that these features would also likely be affected by the acoustic conditions of the signal. Pitch features estimated as F0 values are affected by background noise which produces halving or doubling errors at varying rates. Further, speech rate or rhythm features extracted from automatic alignments would also be affected by acoustic conditions since the quality of these alignments greatly depends on those conditions. Manual phone- or syllable-level alignments, though, would allow for the computation of robust speech rate or rhythm features that can be reliably used even on datasets with unreliable acoustic conditions.

\vspace{-0.2cm}
\section{Conclusions}
\vspace{-0.1cm}

We studied the performance of two acoustic systems for identifying AD patients from healthy controls on the ADreSS$_o$ dataset and on a proprietary dataset in Spanish. We showed that, in both cases, a system run only on non-speech regions leads to above-chance results. That is, in these datasets, the class of the sample (AD or control) can be partially predicted using only the acoustic conditions of the signal, without access to speech information. This is likely due to a  difference in the acquisition conditions across groups. Given this worrisome result, we recommend that datasets collected without a strict control of acoustic conditions should not be used for studies involving acoustic features. Even if analyses are restricted  to the signals’ speech portions, bias in acoustic conditions may  confound the results.

\vspace{-0.2cm}
\section{Acknowledgements}
\vspace{-0.1cm}
\footnotesize
We gratefully acknowledge the support of NVIDIA Corporation for the donation of a Titan Xp GPU. Adolfo García is supported by GBHI, Alzheimer’s Association, and Alzheimer’s Society (Alzheimer’s Association GBHI ALZ UK-22-865742); and ANID (FONDECYT Regular 1210176).

\bibliographystyle{IEEEtran}

\bibliography{template}
\end{document}